\documentclass[12pt]{article}

\usepackage{fancyhdr}
\usepackage{fancybox}
\usepackage{a4wide}
\usepackage{amsmath,amsfonts,amsthm,amssymb}
\usepackage[dvips]{graphics}
\usepackage[pdftex]{graphicx}
\usepackage{epsfig}
\usepackage{epsf}

\newtheorem{thm}{Theorem}[subsection]
 \newtheorem{cor}[thm]{Corollary}
 \newtheorem{lem}[thm]{Lemma}
 \newtheorem{prop}[thm]{Proposition}
 \newtheorem{defn}[thm]{Definition}
 \newtheorem{rem}[thm]{Remark}
 
 \newtheorem{ass}[thm]{Assumptions}

\begin{document}
\title{On EM algorithms and their proximal generalizations}

\author{St\'ephane Chr\'etien and Alfred O. Hero}

\maketitle
%
%
\footnotetext[1]{Stephane Chretien is with Universit\'{e} de Franche-Comt\'e,
 Laboratoire de Math\'ematiques, UMR CNRS 6623, 16 route de Gray, 25030 Besan{\c c}on
(chretien@math.univ-fcomte.fr) and Alfred Hero is with the Dept.  of Electrical
  Engineering and Computer Science, 1301 Beal St., University of Michigan,
  Ann Arbor, MI 48109-2122 (hero@eecs.umich.edu).  This research was
  supported in part by AFOSR grant F49620-97-0028}

\begin{abstract} In this paper, we analyze the celebrated EM algorithm from 
the point of view of proximal point algorithms. More precisely, we study a new type of generalization of the EM 
procedure introduced in \cite{Chretien&Hero:98} and called Kullback-proximal algorithms. 
The proximal framework allows us to prove new 
results concerning the cluster points. An essential contribution is a detailed analysis of the case where 
some cluster points lie on the boundary of the parameter space.
\end{abstract}
%
%
%
%


\section{Introduction}

The problem of maximum likelihood (ML) estimation consists of finding
a solution of the form
\begin{equation}
\label{ml}
\theta_{ML} = {\rm argmax}_{\theta \in \Theta} \; l_y(\theta),
\end{equation}
where $y$ is an observed sample of a random variable $Y$ defined on a
sample space $\mathcal Y$ and $l_y(\theta)$ is the log-likelihood
function defined by
\begin{equation}
l_y(\theta)=\log g(y;\theta),
\end{equation}
defined on the parameter space $\Theta\subset \mathbb R^n$,
and $g(y;\theta)$ denotes the density of $Y$ at $y$ parametrized by the vector
parameter $\theta$.

The Expectation Maximization (EM) algorithm is an iterative procedure which
is widely used for solving ML estimation problems. The EM algorithm
was first proposed by Dempster, Laird and Rubin \cite{Dempster&etal:JRSS77}
and has seen the number of its potential applications increase substantially since
its appearance. The book of McLachlan and Krishnan \cite{McLachlan&Krishnan:97}
gives a comprehensive overview of the theoretical properties of the method and its
applicability.

The convergence of the sequence of EM iterates towards a maximizer of the
likelihood function was claimed in the original paper \cite{Dempster&etal:JRSS77}
but it was later noticed that the proof contained a flaw. A careful convergence analysis
was finally given by Wu \cite{Wu:83} based on Zangwill's general theory \cite{Zangwill:69};
see also \cite{McLachlan&Krishnan:97}. Zangwill's theory applies to general iterative schemes and the main
task when using it is to verify that the assumptions of Zangwill's theorems are satisfied.
Since the appearance of Wu's paper, convergence of the EM algorithm is often taken for granted in many cases where
the necessary assumptions were sometimes not carefully justified. As an example, an often neglected
issue is the behavior of EM iterates when they approach the boundary of the domain of definition
of the functions involved. A different example is the following. It is natural to try and establish that EM iterates
actually converge to a single point $\theta^*$, which involves proving uniqueness of the cluster point. Wu's approach, reported in \cite[Theorem 3.4, p. 89]{McLachlan&Krishnan:97} is based on the assumption that the euclidean distance between two successive iterates
tends to zero. However such an assumption is in fact very hard to verify in most cases and should not be deduced solely from 
experimental observations.

The goal of the present paper is to propose an analysis of EM iterates and their generalizations in the
framework of Kullback proximal point algorithms. We focus on the geometric conditions that are 
provable in practice and the concrete difficulties concerning convergence towards boundaries and
cluster point uniqueness. The approach adopted here was first proposed in \cite{Chretien&Hero:98} in which it was shown
that the EM algorithm could be recast as a Proximal Point algorithm. 
A proximal scheme for maximizing the function $l_y(\theta)$ using the distance-like function $I_y$ is
an iterative procedure of the form
\begin{equation}
\label{proxdef}
\theta^{k+1} \in {\rm argmax}_{\theta \in \Omega} l_y(\theta)-\beta_k I_y(\theta, \theta^k),
\end{equation}
where $(\beta_k)_{k\in \mathbb N}$ is a sequence of positive real numbers often called relaxation parameters.
Proximal point methods were introduced by Martinet \cite{Martinet:70} and
Rockafellar \cite{Rockafellar:76} in the context of convex minimization.
The proximal point representation of the EM algorithm \cite{Chretien&Hero:98}
is obtained by setting $\beta_k=1$ and $I_y(\theta,\theta^k)$ to the Kullback distance between some well specified conditional densities of
a complete data vector. The general case of $\beta_k>0$ was called the Kullback Proximal Point
algorithm (KPP). This approach was further developed in
\cite{Chretien&Hero:00} where  convergence was studied in the twice differentiable case with the assumption that the limit 
point lies in the interior of the domain. The main novelty of \cite{Chretien&Hero:00} was to prove that relaxation of the Kullback-type penalty could ensure 
superlinear convergence which was confirmed by experiment for a Poisson linear inverse problem. This paper is 
an extension of these previous works that addresses the problem of convergence under general conditions. 

The main results of this paper are the following. Firstly, we prove that all the cluster points of the Kullback proximal sequence
which lie in the interior of the domain are stationary points of the likelihood function $l_y$ under
very mild assumptions that are easily verified in practice. Secondly, taking into account finer properties of $I_y$, 
we prove that every cluster point on the boundary of the domain satisfies the Karush-Kuhn-Tucker necessary conditions for optimality under nonnegativity constraints. To illustrate our results, we apply the Kullback-proximal algorithm 
to an estimation problem in animal carcinogenicity introduced in \cite{Ahn&al:00} in which an interesting nonconvex constraint is handled. In this case, the M-step cannot be obtained in closed form. However, the Kullback-proximal algorithm can be analyzed and implemented. Numerical experiments are provided which demonstrate the ability of the method to significantly accelerate the convergence of standard EM.  

The paper is organized as follows. In Section \ref{Kull}, we review the Kullback proximal point interpretation of 
EM. Then, in Section \ref{conv}
we study the properties of interior cluster points. We prove that such cluster points are in fact global maximizers of a certain
penalized likelihood function. This allows us to justify using a relaxation parameter $\beta$ when $\beta$ is 
sufficiently small to permit avoiding saddle points. 
Section \ref{conv2} pursues the analysis in the case where the cluster point lies on a boundary of the domain of $I_y$. 

\section{The Kullback proximal framework}
\label{Kull}
In this section, we review the EM algorithm and the Kullback proximal
interpretation discussed in \cite{Chretien&Hero:00}.

\subsection{The EM algorithm}

The EM procedure is an iterative method which produces a sequence
$(\theta^k)_{k\in \mathbb N}$ such that each $\theta^{k+1}$ maximizes a local
approximation of the likelihood function in the neighborhood of $\theta^k$. This
point of view will become clear in the proximal point framework of the next subsection.

In the traditional approach, one assumes that some data are hidden from the observer. A frequent
example of hidden data is the class to which each sample belongs in the case of mixtures
estimation. Another example is when the observed data are projection of an unkown object
as for image reconstruction problems in tomography. One would prefer to consider the likelihood
of the complete data instead of the ordinary likelihood. Since some parts of the data are hidden,
the so called complete likelihood cannot be computed and therefore must be
approximated. For this purpose, we will need some appropriate notations and assumptions
which we now describe. The observed data are assumed to be i.i.d. samples from a
unique random vector $Y$ taking values on a data space $\mathcal Y$.
Imagine that we have at our disposal more informative
data than just samples from $Y$. Suppose that the more informative data are
samples from a random variable $X$ taking values on a space $\mathcal X$ with density
$f(x;\theta)$ also parametrized by $\theta$.  We will say that the data $X$ is more informative
than the actual data $Y$ in the sense that $Y$ is a compression of $X$, i.e. there exists
a non-invertible transformation $h$ such that $Y=h(X)$.  If one had access to the data $X$ it
would therefore be advantageous to replace the ML estimation problem
(\ref{ml}) by
\begin{equation}
\label{mlx}
\hat{\theta}_{ML}={\rm argmax}_{\theta \in {\mathbb R}^p} l_x(\theta),
\end{equation}
with $l_x(\theta)=\log f(x;\theta)$.
Since $y=h(x)$ the density $g$ of $Y$ is related to the density $f$ of $X$
through
\begin{equation}
\label{cond}
g(y;\theta)=\int_{h^{-1}(\{y\})}f(x;\theta)d\mu(x)
\end{equation}
for an appropriate measure $\mu$ on $\mathcal X$.
In this setting, the data $y$ are called {\em
incomplete data} whereas the data $x$ are called {\em complete data}.

Of course the complete data $x$ corresponding to a given observed sample $y$
are unknown. Therefore, the complete data likelihood function $l_x(\theta)$
can only be estimated. Given the observed data $y$ and a previous estimate of
$\theta$ denoted $\bar{\theta}$, the following minimum mean square error
estimator (MMSE) of the quantity $l_x(\theta)$ is natural
\begin{equation*}
Q(\theta,\bar{\theta})={\sf E}[\log f(x;\theta)| y;\bar{\theta}],
\end{equation*}
where, for any integrable function $F(x)$ on $\mathcal X$, we have
defined the conditional expectation
\begin{equation*}
{\sf E}[F(x)| y;\bar{\theta}]=\int_{h^{-1}(\{y\})}
F(x) k(x| y;\bar{\theta}) d\mu(x)
\end{equation*}
and $k(x| y;\bar{\theta})$ is the conditional density function given $y$
\begin{equation}\label{condi}
k(x| y;\bar{\theta})=\frac{f(x;\bar{\theta})}{g(y;\bar{\theta})}.
\end{equation}

Having described the notions of complete data and complete likelihood and its
local estimation we now turn to the EM algorithm. The idea is relatively
simple: a legitimate way to proceed is to require that iterate $\theta^{k+1}$
be a maximizer of the local estimator of the complete likelihood conditionally
on $y$ and $\theta^k$. Hence, the EM algorithm generates a sequence of
approximations to the solution (\ref{mlx}) starting from an initial guess
$\theta^0$ of $\theta_{ML}$ and is defined by
\begin{equation*}
\text{\bf Compute } Q(\theta,\theta^k)={\sf E}[\log f(x;\theta)| y;\theta^k]
\text{\hspace{1cm}\bf E Step}\nonumber
\end{equation*}
\begin{equation*}
\theta^{k+1}={\rm argmax}_{\theta \in {\mathbb R}^p} Q(\theta,\theta^k)
\text{\hspace{2.9cm}\bf M Step}\nonumber
\end{equation*}

\subsection{Kullback proximal interpretation of the EM algorithm}
\label{proxEM}
Consider the general problem of maximizing a concave function $\Phi(\theta)$.
The original proximal point algorithm introduced by Martinet \cite{Martinet:70}
is an iterative procedure which can be written
\begin{equation}
\label{proxit}
\theta^{k+1}={\rm argmax}_{\theta \in D_\Phi}\left\{\Phi(\theta)
-\frac{\beta_k}{2} \|\theta-\theta^k\|^2\right\}.
\end{equation}
The quadratic penalty $\frac12 \|\theta-\theta^k\|^2$ is relaxed using a
sequence of positive parameters $\{\beta_k\}$. In
\cite{Rockafellar:76}, Rockafellar showed that superlinear
convergence of this method is obtained when the sequence $\{\beta_k\}$
converges towards zero.

It was proved in \cite{Chretien&Hero:00} that the EM algorithm is a particular example in the class
of proximal point algorithms using Kullback Leibler types of penalties. One
proceeds as follows. Assume that the family of conditional densities $\{k(x|
y;\theta)\}_{\theta \in {\mathbb R}^p}$ is regular in the sense of Ibragimov
and Khasminskii \cite{Ibragimov&Has'minskii:81}, in particular $k(x|
y;\theta)\mu(x)$ and $k(x| y;\bar{\theta)}\mu(x)$ are mutually absolutely
continuous for any $\theta$ and $\bar{\theta}$ in ${\mathbb R}^p$. Then the
Radon-Nikodym derivative $\frac{k(x| y,\bar{\theta})}{k(x| y;\theta)}$ exists
for all $\theta,\bar{\theta}$ and we can define the following Kullback Leibler divergence:
\begin{equation}
\label{kullb}
I_y(\theta,\bar{\theta})={\sf E}\bigl[
\log \frac{k(x| y,\bar{\theta})}{k(x| y;\theta)}| y;\bar{\theta} \;
\bigr].
\end{equation}
We are now able to define the Kullback-proximal algorithm. For this purpose, let us
define $D_l$ as the domain of $l_y$, $D_{I,\theta}$ the domain of $I_y(\cdot,\theta)$ and
$D_I$ the domain of $I_y(\cdot,\cdot)$.
\begin{defn}
Let $(\beta_k)_{k\in \mathbb N}$ be a sequence of positive real numbers.
Then, the Kullback-proximal algorithm is defined by
\begin{equation}
\label{kullprox}
\theta^{k+1}={\rm argmax}_{\theta\in D_l\cap D_{I,\theta^k}}
l_y(\theta)- \beta_k I_y(\theta,\theta^k).
\end{equation}
\end{defn}
The main result on which the present paper relies is that EM algorithm is a
special case of (\ref{kullprox}), i.e. it is a penalized ML estimator with proximal penalty 
$I_y(\theta,\theta^k)$.

\begin{prop}{\rm \cite[Proposition 1]{Chretien&Hero:00}}
\label{equiem}
The EM algorithm is a special instance of the Kullback-proximal algorithm with $\beta_k =1$,
for all $k\in \mathbb N$.
\end{prop}

The previous definition of the Kullback proximal algorithm may appear overly general to 
the reader familiar with the usual practical interpretation of the EM algorithm. However, we found that 
such a framework has at least the three following benefits \cite{Chretien&Hero:00}: 
\begin{itemize}
\item to our opinion, the convergence proof of our EM is more natural,
\item the Kullback proximal framework may easily incorporate additional constraints, a feature that 
may be of crucial importance as demonstrated in the example of Section \ref{prob2} below,
\item the relaxation sequence $(\beta_k)_{k\in \mathbb N}$ allows one to weight the penalization
term and its convergence to zero implies quadratic convergence in certain 
examples.
\end{itemize}

The first of these three arguments is also supported by our simplified treatment of the componentwise 
EM procedure proposed in \cite{Celeux&etal:01} and the remarkable recent results of \cite{Tseng:04} 
based on a special proximal entropic representation of EM for getting precise estimates on the convergence speed of EM algorithms, however, with much more restrictive assumptions than the ones of the present paper. 

Although our results are obtained 
under mild assumptions concerning the relaxation sequence $(\beta_k)_{k\in \mathbb N}$ including the 
case $\beta_k=0$, several precautions should be taken when implementing the method. However, 
one of the key features of EM-like procedures is to allow easy 
handling of positivity or more complex constraints, such as the ones discussed in the example of 
Section \ref{prob2}. In such cases the function $I_y$ behaves like a barrier whose value 
increases to infinity as the iterates approach the boundary of the constraint set. 
Hence, the sequence $(\beta_k)_{k\in \mathbb N}$ ought to be positive in 
order to exploit this important computational feature. On the other hand, 
as proved under twice differentiability assumptions in \cite{Chretien&Hero:00} when the 
cluster set reduces to a unique nondegenerate maximizer in the interior of the domain of 
the log-likelihood and $\beta_k$ converges to zero, quadratic convergence is obtained. This 
nice behavior is not satisfied in the plain EM case where $\beta_k=1$ for all $k\in \mathbb N$. 
As a drawback, one problem in decreasing the $\beta_k$'s too quickly is possible 
numerical ill conditioning. The problem of choosing the relaxation sequence 
is still largely open. We have found however that for most "reasonable" sequences, our method 
was at least as fast as the standard EM. 

Finally, we would like to end our presentation of KPP-EM by noting that closed form iterations may not be 
available in the case $\beta_k\neq 1$. If this is the case,  solving (\ref{kullprox}) becomes a subproblem which will
require iterative algorithms. In some interesting examples, e.g. the case presented in Section 
\ref{prob2}. In this case, the standard EM iterations are not available 
in closed form in the first place and KPP-EM provides faster convergence while preserving monotonicity and constraint satisfaction.

\subsection{Notations and assumptions}

The notation $\|\cdot\|$ will be used to denote the norm on any
previously defined space without more precision. The space on
which it is the norm should be obvious from the context. For any bivariate
function $\Phi$, $\nabla_1\Phi$ will denote the gradient with respect to the first variable.
In the remainder of this paper we will make the following assumptions.
\begin{ass}
\label{ass1}
{\rm (i) $l_y$ is differentiable on $D_l$ and $l_y(\theta)$
tends to $-\infty$ whenever $\|\theta\|$ tends to $+\infty$. \\
(ii) the projection of $D_I$ onto the first coordinate is a subset of $D_l$. \\
(iii) $(\beta_k)_{k\in\mathbb N}$ is a convergent nonnegative sequence of real numbers whose
limit is denoted by $\beta^*$.}
\end{ass}
We will also impose the following assumptions on the distance-like function $I_y$.
\begin{ass}
\label{ass2} {\rm (i) There exists a finite dimensional euclidean
space $S$, a differentiable mapping $t : D_l \mapsto S$ and a
functional $\Psi : D_\Psi \subset S\times S \mapsto \mathbb R$
such that
$$
I_y(\theta,\bar{\theta})=\Psi(t(\theta),t(\bar{\theta})),
$$
where $D_\psi$ denotes the domain of $\Psi$. \\
(ii) For any $\{t^k,t)_{k\in \mathbb N}\}\subset D_\Psi$ there
exists $\rho_t>0$ such that $\lim_{\|t^k-t\|\rightarrow\infty}
I_y(t^k,t)\geq \rho_t$.
Moreover, we assume that $\inf_{t\in M}\rho_t>0$ for any bounded set $M\subset S$.  \\
For all $(t^\prime,t)$ in $D_\Psi$, we will also require that \\
(iii) (Positivity) $\Psi(t^\prime,t)\geq 0$, \\
(iv) (Identifiability) $\Psi(t^{\prime},t)=0 \Leftrightarrow t=t^\prime$,\\
(v) (Continuity) $\Psi$ is continuous at $(t^{\prime},t)$ \\
and for all $t$ belonging to the projection of $D_\Psi$ onto its
second coordinate,\\
(vi) (Differentiability) the function $\Psi(\cdot,t)$ is
differentiable at $t$.}
\end{ass}

Assumptions \ref{ass1}(i) and (ii) on $l_y$ are standard and are easily checked
in practical examples, e.g. they are satisfied for the Poisson and additive mixture models. 
Notice that the domain $D_I$ is now
implicitly defined by the knowledge of $D_l$ and $D_\Psi$.
Moreover $I_y$ is continuous on $D_I$. The importance of requiring
that $I_y$ has the prescribed shape comes from the fact that $I_y$
might not satisfy assumption \ref{ass2}(iv) in general. Therefore assumption \ref{ass2}
(iv) reflects the requirement that $I_y$ should at least satisfy the
identifiability property up to a possibly injective
transformation. In both examples discussed above, this property is
an easy consequence of the well known fact that $a\log(a/b)=0$
implies $a=b$ for positive real numbers $a$ and $b$. The growth,
continuity and differentiability properties \ref{ass2} (ii), (v)
and (vi) are, in any case, nonrestrictive.

For the sake of notational convenience, the regularized objective
function with relaxation parameter $\beta$ will be denoted
\begin{equation}\label{regfunc}
F_\beta (\theta,\bar{\theta})=l_y(\theta)-\beta
I_y(\theta,\bar{\theta}).
\end{equation}

Finally we make the following general assumption.

\begin{ass}
\label{ass3}
The Kullback proximal iteration (\ref{kullprox}) is well
defined, i.e. there exists at least one maximizer of $F_{\beta^k}
(\theta,\theta^{k})$ at each iteration $k$.
\end{ass}
In the EM case, i.e. $\beta=1$, this last assumption is equivalent to the 
computability of M-steps. A sufficient condition for this assumption to hold would be, for instance, that
$F_\beta(\theta,\bar{\theta})$ be sup-compact, i.e. the level sets
$\{\theta \mid F_\beta(\theta,\bar{\theta})\geq \alpha \}$ be
compact for all $\alpha$, $\beta>0$ and $\bar{\theta}\in D_l$.
However, this assumption is not usually satisfied since the
distance-like function is not defined on the boundary of its
domain. In practice it suffices to solve the equation $\nabla
F_{\beta^k}(\theta,\theta^k)=0$, to prove that the solution is
unique. Then assumption \ref{ass1}(i) is sufficient to conclude
that we actually have a maximizer.

\subsection{General properties : monotonicity and boundedness}

Using Assumptions \ref{ass1}, we easily deduce monotonicity of the likelihood values and boundedness of the
proximal sequence. The first two lemmas are proved, for instance, in \cite{Chretien&Hero:00}.

We start with the following monotonicity result.
\begin{lem}\cite[Proposition 2]{Chretien&Hero:00}
\label{truit} For any iteration $k \in \mathbb N$, the sequence $(\theta^k)_{k\in \mathbb N}$ satisfies
\begin{equation}
\label{majgrad} l_y(\theta^{k+1})-l_y(\theta^k)\geq \beta_k I_y(\theta^k,\theta^{k+1})\geq 0.
\end{equation}
\end{lem}

From the previous lemma, we easily obtain the boundedness of the sequence. 
\begin{lem}\cite[Lemma 2]{Chretien&Hero:00}
\label{boundu}
The sequence $(\theta^k)_{k\in \mathbb N}$ is bounded.
\end{lem}

The next lemma will also be useful.
\begin{lem}
\label{yal} Assume that there exists a subsequence
$(\theta^{\sigma(k)})_{k\in \mathbb N}$ belonging to a compact set
$C$ included in $D_l$. Then,
\begin{equation}
\lim_{k\rightarrow\infty} \beta_{k}
I_y(\theta^{k+1},\theta^{k})=0.
\nonumber
\end{equation}
\end{lem}

{\bf Proof}. Since $l_y$ is continuous over $C$, $\sup_{\theta \in
C}l_y(\theta)<+\infty$ and $(l_y(\theta^{\sigma(k)}))_{k\in
\mathbb N}$ is therefore bounded from above. Moreover, Lemma \ref{truit}
implies that the sequence $(l_y(\theta^{k}))_{k\in \mathbb N}$ is
monotone nondecreasing. Therefore, the whole sequence
$(l_y(\theta^{k}))_{k\in \mathbb N}$ is bounded from above and
convergent. This implies that $\lim_{k\rightarrow\infty}l_y(\theta^{k+1})-l_y(\theta^k)=0$.
Applying Lemma \ref{truit} again, we obtain the desired result.
\hfill$\Box$

\section{Analysis of interior cluster points}\label{conv}
The convergence analysis of Kullback proximal algorithms is split
into two parts, the first part being the subject of this
section. We prove that if the accumulation points $\theta^*$ of the Kullback proximal sequence
satisfy $(\theta^*,\theta^*)\in D_{I_y}$ they are stationary
points of the log-likelihood function $l_y$. It is also straightforward to show that 
the same analysis applies to the case of penalized likelihood estimation. 

\subsection{Nondegeneracy of the Kullback penalization}
We start with the following useful lemma.
\begin{lem}
\label{asymreg} Let $(\alpha_1^k)_{k \in \mathbb N}$ and
$(\alpha_2^k)_{k \in \mathbb N}$ be two bounded sequences in
$D_\Psi$ satisfying 
$$\lim_{k\rightarrow \infty}\Psi(\alpha_1^k,\alpha_2^k)=0.$$ 
Assume that every couple
$(\alpha_1^*,\alpha_2^*)$ of accumulation points of these two
sequences lies in $D_\Psi$. Then,
\begin{equation}
\lim_{k\rightarrow \infty}\|\alpha_1^k-\alpha_2^k\|=0. \nonumber
\end{equation}
\end{lem}

{\bf Proof}. First, one easily obtains that $(\alpha_2^k)_{k \in
\mathbb N}$ is bounded (use a contradiction argument and
Assumption \ref{ass2} (ii)). Assume that there exits a subsequence
$(\alpha_1^{\sigma(k)})_{k\in \mathbb N}$ such that
$\|\alpha_1^{\sigma(k)}-\alpha_2^{\sigma(k)}\|\geq 3\epsilon$ for
some $\epsilon>0$ and for all large $k$. Since
$(\alpha_1^{\sigma(k)})_{k\in \mathbb N}$ is bounded, one can
extract a convergent subsequence. Thus we may assume without
any loss of generality that $(\alpha_1^{\sigma(k)})_{k\in \mathbb
N}$ is convergent with limit $\alpha^*$. Using the triangle
inequality, we have
$\|\alpha_1^{\sigma(k)}-\alpha_1^*\|+\|\alpha_1^*-\alpha_2^{\sigma(k)}\|\geq
3\epsilon$. Since $(\alpha_1^{\sigma(k)})_{k\in \mathbb N}$
converges to $\alpha_1^*$, there exists a integer $K$ such that
$k\geq K$ implies $\|\alpha_1^{\sigma(k)}-\alpha_1^*\|\leq
\epsilon$. Thus for $k\geq K$ we have
$\|\alpha_1^*-\alpha_2^{\sigma(k)}\|\geq 2\epsilon$. Now recall
that $(\alpha_2^k)_{k \in
  \mathbb N}$ is bounded and extract a convergent subsequence
$(\alpha_2^{\sigma(\gamma(k))})_{k\geq K}$ with limit denoted by
$\alpha_2^*$. Then, using the same arguments as above, we obtain
$\|\alpha_1^*-\alpha_2^*\|\geq \epsilon$. Finally, recall that
$\lim_{k\rightarrow\infty}\Psi(\alpha_1^k,\alpha_2^k)=0$. We thus
have
$\lim_{k\rightarrow\infty}\Psi(\alpha_1^{\sigma(\gamma(k))},\alpha_2^{\sigma(\gamma(k))})=0$,
and, due to the fact that the sequences are bounded and
$\Psi(\cdot,\cdot )$ is continuous in both variables, we have
$I_y(\alpha_1^*,\alpha_2^*)=0$. Thus assumption \ref{ass2} (iv)
implies that $\|\alpha_1^*-\alpha_2^*\|=0$ and we obtain a
contradiction. Hence, $\lim_{k\rightarrow
\infty}\|\alpha_1^k-\alpha_2^k\|=0$ as claimed. \hfill $\Box$

\subsection{Cluster points}
The main results of this section are the following. First, we
prove that under the assumptions \ref{ass1}, \ref{ass2} and \ref{ass3}, any cluster point
$\theta^*$ is a global maximizer of
$F_{\beta^*}(\theta^*,\theta^*)$. We then use this general result
to prove that such cluster points are
stationary points of the log-likelihood function. This result motivates a natural assumption under which $\theta^*$ is
in fact a local maximizer of $l_y$. In addition we show that if the
sequence $(\beta^k)_{k\in \mathbb N}$ converges to zero, i.e.
$\beta^*=0$, then $\theta^*$ is a global maximizer of
log-likelihood. Finally, we discuss some simple conditions under
which the algorithm converges, i.e. has only one cluster point. 

The following theorem states a result which describes the stationary points of the proximal point
algorithm as global maximizers of the asymptotic penalized function.
\begin{thm}
\label{the} Assume that $\beta^*>0$. Let $\theta^*$ be any
accumulation point of $(\theta^k)_{k\in \mathbb N}$. Assume that
$(\theta^*,\theta^*) \in D_I$. Then, $\theta^*$ is a global
maximizer of the penalized function $F_{\beta^*}(\cdot,\theta^*)$
over the projection of $D_I$ onto its first coordinate, i.e. 
\begin{equation}
F_{\beta^*}(\theta^*,\theta^*)\geq F(\theta,\theta^*) 
\nonumber
\end{equation}
for all $\theta$ such that $(\theta,\theta^*)\in D_I$. 
\end{thm}

An informal argument is as follows. Assume that $\Theta=\mathbb R^n$. From the definition of the proximal iterations,
we have
$$
F_{\beta_{\sigma(k)}}(\theta^{\sigma(k)+1},\theta^{\sigma(k)}) \geq
F_{\beta_{\sigma(k)}}(\theta,\theta^{\sigma(k)})
$$
for all subsequence $(\theta^{\sigma(k)})_{k\in \mathbb N}$ converging to $\theta^*$ and
for all $\theta\in \Theta$. Now, assume we can prove that $\theta^{\sigma(k)}$
also converges to $\theta^*$, we obtain by taking the limit and using continuity, that
$$
F_{\beta_*}(\theta^*,\theta^*) \geq
F_{\beta_*}(\theta,\theta^*)
$$
which is the required result. There are two major difficulties when one tries to transform this
sketch into a rigorous argument. The first one is related to the fact that
$l_y$ and $I_y$ are only defined on domains which may not to be closed.
Secondly, proving that $\theta^{\sigma(k)}$ converges to $\theta^*$ is not an easy
task. This issue will be discussed in more detail in the
next section. The following proof overcomes both difficulties.

{\bf Proof}. Without loss of generality, we may reduce the
analysis to the case where $\beta_k\geq \beta>0$ for a certain
$\beta$. The fact that $\theta^*$ is a cluster point implies that
there is a subsequence of $(\theta^{k})_{k\in \mathbb N}$
converging to $\theta^*$. For $k$ sufficiently large, we may assume that the terms
$(\theta^{\sigma(k+1)},\theta^{\sigma(k)})$ belong to a compact neighborhood $C^*$
of $(\theta^*,\theta^*)$ included in $D_I$. Recall that
$$
F_{\beta_{\sigma(k)-1}}(\theta^{\sigma(k)},\theta^{\sigma(k)-1}) \geq
F_{\beta_{\sigma(k)-1}}(\theta,\theta_{\sigma(k)-1})
$$
for all
$\theta$ such that $(\theta,\theta^{\sigma(k)-1})\in D_I$ and a fortiori for
$(\theta,\theta^{\sigma(k)-1})\in C^*$. Therefore,
\begin{equation}
\label{eq0}
\begin{array}{rl}
F_{\beta^*}(\theta^{\sigma(k)},\theta^{\sigma(k)-1})
& -(\beta_k-\beta^*) I_y(\theta^{\sigma(k)},\theta^{\sigma(k)-1}) \geq \\ 
&  F_{\beta^*}(\theta,\theta^{\sigma(k)-1})-(\beta_{\sigma(k)-1}-\beta^*)
I_y(\theta,\theta^{\sigma(k)-1}).
\end{array}
\end{equation}

Let us have a precise look at the "long term" behavior of $I_y$.
First, since $\beta_k>\beta_*$ for all $k$ sufficiently large, Lemma \ref{yal} says that
$$
\lim_{k\rightarrow\infty}I_y(\theta^{\sigma(k)},\theta^{\sigma(k)-1})=0.
$$
Thus, for any $\epsilon>0$, there exits an integer $K_1$ such that
$I_y(\theta^{\sigma(k)},\theta^{\sigma(k)+1})\leq \epsilon$ for
all $k\geq K_1$. Moreover, Lemma \ref{asymreg} and continuity of $t$
allows to conclude that
$$\lim_{k\rightarrow \infty} t(\theta^{\sigma(k)-1})=t(\theta^*).$$
Since $\Psi$ is continuous, for all $\epsilon>0$ and for all
$k$ sufficienlty large we have
\begin{equation}
\label{eq}
\begin{array}{rl}
I_y(\theta^*,\theta^*) & = \Psi(t(\theta^*),t(\theta^*))\\
& \geq \Psi(t(\theta^{\sigma(k)}),t(\theta^{\sigma(k)-1}))-\epsilon \\
& = I_y(\theta^{\sigma(k)},\theta^{\sigma(k)-1}) -\epsilon.
\end{array}
\end{equation}

On the other hand, $F_{\beta^*}$ is continuous
in both variables on $C^*$, due to Assumptions \ref{ass1}(i) and
\ref{ass2}(i). By continuity in the first
and second arguments of $F_{\beta^*}(\cdot,\cdot )$, for any
$\epsilon > 0$ there exists $K_2\in \mathbb N$ such that for all $k\geq K_2$
\begin{equation}\label{eq1}
F_{\beta^*}(\theta^*,\theta)\leq F_{\beta^*}(\theta^{\sigma(k)},\theta)+\epsilon.
\end{equation}
Using (\ref{eq}), since $l_y$ is continuous, we obtain existence
of $K_3$ such that for all $k\geq K_3$
\begin{equation}\label{eq2}
F_{\beta^*}(\theta^*,\theta^*)\geq
F_{\beta^*}(\theta^{\sigma(k)},\theta^{\sigma(k)+1})-2\epsilon.
\end{equation}
Combining equations (\ref{eq1}) and (\ref{eq2}) with (\ref{eq0}), we obtain
\begin{equation}
\label{tronte}
\begin{array}{rl}
F_{\beta^*}(\theta^*,\theta^*)\geq & F_{\beta^*}(\theta^*,\theta)-(\beta_k-\beta^*)
I_y(\theta^{\sigma(k)},\theta) \\
& \hspace{.5cm} +(\beta_k-\beta^*)I_y(\theta^{\sigma(k)},\theta^{\sigma(k)+1})) -3\epsilon.
\end{array}
\end{equation}
Now, since $\beta^*=\lim_{k\rightarrow \infty}\beta_k$, there
exists an integer $K_4$ such that $\beta_k-\beta^*\leq \epsilon$ for
all $k\geq K_4$. Therefore for all $k\geq \max \{K_1,K_2,K_3,K_4\}$, we obtain
\begin{equation*}
\begin{array}{rl}
F_{\beta^*}(\theta^*,\theta^*)\geq F_{\beta^*}(\theta^*,\theta) -\epsilon I_y(\theta^{\sigma(k)},\theta)-\epsilon^2-3\epsilon.
\end{array}
\end{equation*}
Since $I_y$ is continuous and $(\theta^{\sigma(k)})_{k\in\mathbb N}$ is
bounded, there exists a real constant $K$ such that $I_y(\theta^{\sigma(k)},\theta)\leq
K$, for all $n\in \mathbb N$. Thus, for all $k$ sufficiently large
\begin{equation}
\label{maj} F_{\beta^*}(\theta^*,\theta^*)\geq
F_{\beta^*}(\theta^*,\theta)-(4\epsilon K+\epsilon^2).
\end{equation}
Finally, recall that no assumption was made on $\theta$, and that
$C^*$ is any compact neighborhood of $\theta^*$. Thus, using the
assumption \ref{ass1}(i), which asserts that $l_y(\theta)$ tends to
$-\infty$ as $\|\theta\|$ tends to $+\infty$, we may deduce that
(\ref{maj}) holds for any $\theta$ such that $(\theta,\theta^*)
\in D_I$ and, letting $\epsilon$ tend to zero, we see that
$\theta^*$ maximizes $F_{\beta^*}(\theta,\theta^*)$ for over all
$\theta$ such that $(\theta,\theta^*)$ belongs to $D_I$ as
claimed. \hfill$\Box$

Using this theorem, we may now deduce
that certain accumulation points on the strict interior of the
parameter's space are stationary points of the log-likelihood
function.

\begin{cor}
\label{stat}
\label{corolun} Assume that $\beta^*>0$. Let $\theta^*$ be any
accumulation point of $(\theta^k)_{k\in \mathbb N}$. Assume that
$(\theta^*,\theta^*) \in {\rm int} D_I$. Then, if $l_y$ is
differentiable on $D_l$, $\theta^*$
is a stationary point of $l_y(\theta)$. Moreover, if $l_y$ is
concave, then $\theta^*$ is a global maximizer of $l_y$.
\end{cor}

{\bf Proof}. Since under the required assumptions $l_y$ is
differentiable and $I_y(\theta^*,\cdot)$ is differentiable at
$\theta^*$, Theorem \ref{the} states that
\begin{equation}
0\in \Big\{ \nabla l_y(\theta^*)+\beta_* \nabla I_y(\theta^*,\theta^*)\Big\}.
\nonumber
\end{equation}
Since $I_y(\cdot,\theta^*)$ is minimum at $\theta^*$, $\nabla_1 I_y(\theta^*,\theta^*)=0$
and we thus obtain that $\theta^*$ is a stationary point of $l_y$. This implies that $\theta^*$ is a global maximizer in the case where $l_y$ is concave.
\hfill$\Box$.

Theorem \ref{the} seems to be much stronger than the previous corollary.
The fact that accumulation points of the proximal sequence may not be global
maximizers of the likelihood is now easily seen to be a consequence of fact that
the Kullback distance-like function $I_y$ perturbs the shape of the likelihood
function when $\theta$ is far from $\theta^*$. This perturbation does not have
serious consequence in the concave case. On the other hand, one may wonder
whether $\theta^*$ cannot be proved to be at least a local maximizer instead
of a mere stationary point. The answer is given in the following corollary.

\begin{cor}
Let $\theta^*$ be an accumulation point of $(\theta^k)_{k\in
\mathbb N}$ such that $(\theta^*,\theta^*) \in{\rm int} D_I$. In
addition, assume that $l_y$ and $I_y(\cdot,\theta^*)$ are twice
differentiable in a neighborhood of $\theta^*$ and that the
Hessian matrix $\nabla^2 l_y(\theta^*)$ at $\theta^*$ is not the
null matrix. Then, if $\beta^*$ is sufficiently small, $\theta^*$
is a local maximizer of $l_y$ over $D_l$.
\end{cor}

{\bf Proof}. Assume that $\theta^*$ is not a local maximizer.
Since $\nabla^2 l_y$ is not the null matrix, for $\beta^*$
sufficiently small, there is a direction $\delta$ in the tangent
space to $D_l$ for which the function
$f(t)=F_{\beta^*}(\theta^*+t\delta,\theta^*)$ has positive second
derivative for $t$ sufficiently small. This contradicts the fact
that $\theta^*$ is a global maximizer of
$F_{\beta^*}(\cdot,\theta^*)$ and the proof is completed.
\hfill$\Box$

The next theorem establishes global optimality
of accumulation points in the case where the relaxation sequence converges to zero.
\begin{thm}
Let $\theta^*$ be any accumulation point of $(\theta^k)_{k\in
\mathbb N}$. Assume that $(\theta^*,\theta^*) \in D_I$.  Then,
without assuming differentiability of either $l_y$ or of $I_y$, if
$(\beta_k)_{k\in\mathbb N}$ converges to zero, $\theta^*$ is a
global maximizer of $l_y$ over the projection of $D_I$ along the
first coordinate.
\end{thm}

{\bf Proof}. Let $(\theta^{\sigma(k)})_{k\in\mathbb N}$ be a
convergent subsequence of $(\theta^k)_{k\in\mathbb N}$ with limit
denoted $\theta^*$. We may assume that for $k$ sufficiently large,
$(\theta^{\sigma(k+1)},\theta^{\sigma(k)})$ belongs to a compact
neighborhood $C^*$ of $\theta^*$. By continuity of $l_y$, for any
$\epsilon>0$, there exists $K\in\mathbb N$ such that for all
$k\geq K$,
\begin{equation}
l_y(\theta^*)\geq l_y(\theta^{\sigma(k)})-\epsilon.
\nonumber\end{equation}
On the other hand, the proximal iteration (\ref{proxdef}) implies that
\begin{equation}
l_y(\theta^{\sigma(k)})-\beta_{\sigma(k)-1}I_y(\theta^{\sigma(k)-1},\theta^{\sigma(k)})
\geq l_y(\theta)-\beta_{\sigma(k)-1}I_y(\theta^{\sigma(k)-1},\theta),
\nonumber\end{equation}
for all $\theta\in D_l$. Fix $\theta\in D_l$. Thus, for all $k\geq K$,
\begin{equation}
l_y(\theta^*)\geq l_y(\theta)+\beta_{\sigma(k)-1}I_y(\theta^{\sigma(k)-1},\theta^{\sigma(k)})
-\beta_{\sigma(k)-1}I_y(\theta^{\sigma(k)-1},\theta)-\epsilon.
\nonumber\end{equation}
Since $I_y$ is a nonnegative function and $(\beta_k)_{k\in \mathbb N}$ is
a nonnegative sequence, we obtain
\begin{equation}
l_y(\theta^*)\geq l_y(\theta)-\beta_{\sigma(k)-1}I_y(\theta^{\sigma(k)-1},\theta)-\epsilon.
\nonumber\end{equation}
Recall that $(\theta^k)_{k\in\mathbb N}$ is bounded due to Lemma \ref{boundu}. Thus,
since $I_y$ is continuous, there exists a constant $C$ such that
$I_y(\theta^{\sigma(k)-1},\theta)\leq C$ for all $k$. Therefore, for $k$ greater than $K$,
\begin{equation}
l_y(\theta^*)\geq l_y(\theta)-\beta_{\sigma(k)-1}C-\epsilon.
\nonumber\end{equation}
Passing to the limit, and recalling that $(\beta_k)_{k\in\mathbb N}$ tends
to zero, we obtain that
\begin{equation}
l_y(\theta^*)\leq l_y(\theta)-\epsilon. \nonumber\end{equation}
Using the same argument as at the end of the proof of Theorem
\ref{the}, this latter equation holds for any $\theta$ such that
$(\theta,\theta^*)$ belongs to $D_I$, which concludes the proof
upon letting $\epsilon$ tend to zero. \hfill$\Box$

\subsection{Convergence of the Kullback proximal sequence}
One question remains open in the analysis of the previous section: does the sequence
generated by the Kullback proximal point converge? In other words: are there multiple
cluster points? In Wu's paper \cite{Wu:83}, the answer takes the following form. If the euclidean distance between two successive
iterates tends to zero, a well known result states that the set of accumulation points is
a continuum (see for instance \cite[Theorem 28.1]{Ostrowski:66}) and therefore, it is connected.
Therefore, if the set of stationary points of $l_y$ is a countable set, the
iterates must converge. 

\begin{thm}
\label{corle} Let $S^*$ denote the set of accumulation points of the sequence $(\theta^k)_{k\in \mathbb N}$.
Assume that $\lim_{k\rightarrow\infty}\|\theta^{k+1}-\theta^k\|=0$
and that $l_y(\theta)$ is strictly concave in an open neighborhood
$\mathcal N$ of an  accumulation point $\theta^*$ of $(\theta^k)_{k\in
\mathbb N}$ and that $(\theta^*,\theta^*)$ is in ${\rm int} D_I$. Then, for any relaxation sequence $(\beta_k)_{k\in \mathbb N}$, the sequence
$(\theta^k)_{k\in \mathbb N}$ converges to a local maximizer of $l_y(\theta)$.
\end{thm}

{\bf Proof}. We obtained in Corollary \ref{stat} that every
accumulation point $\theta^*$ of $(\theta^k)_{k\in \mathbb N}$ in ${\rm int} D_{l_y}$ and such that
$(\theta^*,\theta^*) \in {\rm int} D_{I_y}$ is a stationary
point of $l_y(\theta)$. Since $l_y(\theta)$ is strictly concave over $\mathcal
N$, the set of stationary points of $l_y$ belonging to $\mathcal
N$ reduces to singleton. Thus $\theta^*$ is the unique stationary point
in $\mathcal N$ of $l_y$, and {\em a fortiori}, the unique
accumulation point of $(\theta^k)_{k\in \mathbb N}$ belonging to
$\mathcal N$. To complete the proof, it remains to show that there
is no accumulation point in the exterior of  $\mathcal N$. For
that purpose, consider an open ball $\mathcal B$ of center $\theta^*$
and radius $\epsilon$ included in $\mathcal N$. Then, $x^*$ is the
unique accumulation point in $\mathcal B$.  Moreover, any
accumulation point $\theta^{\prime}$, lying in the exterior of
$\mathcal N$ must satisfy $\|\theta^*-\theta^{\prime}\|\geq \epsilon$, and
we obtain a contradiction with the fact that $S^*$ is connected.
Thus every accumulation point lies in $\mathcal N$, from which we
conclude that $\theta^*$ is the only accumulation point of $(\theta^k)_{k\in
\mathbb N}$  or, in other words, that $(\theta^k)_{k\in \mathbb N}$
converges towards $\theta^*$.  Finally, notice that the strict
concavity of $l_y(\theta)$ over $\mathcal N$ implies  that $\theta^*$ is a
local maximizer. \hfill $\Box$

Before concluding this section, let us make two general remarks.
\begin{itemize}
\item Proving {\em a priori} that the set of stationary points of $l_y$ is discrete may be a hard task
in specific examples.

\item In general, it is not known whether $\lim_{k\rightarrow\infty}\|\theta^{k+1}-\theta^k\|=0$
holds. In fact, Lemma \ref{asymreg} could be a first step in this direction. Indeed if we
could prove in any application that the mapping $t$ is injective, the
desired result would follow immediately. However, injectivity of $t$ does not hold
in many of the standard examples; in the case of Gaussian mixtures, 
see \cite[Section 2.2]{Celeux&etal:01} for instance. Thus we are now able
to clearly understand why the assumption that $\lim_{k\rightarrow\infty}\|\theta^{k+1}-\theta^k\|=0$
is not easily deduced from general arguments. This
problem has been overcome in \cite{Celeux&etal:01} where it is shown that $t$ is
componentwise injective and thus performing a componentwise EM algorithm is a 
good alternative to the standard EM.
\end{itemize}

\section{Analysis of cluster points on the boundary}\label{conv2}
The goal of this section is to extend the previous results to the case where some cluster points lie on
the boundary of the region where computation of proximal steps is well defined. Such cluster points have
rarely been analyzed in the statistical literature and the strategy developed for the interior case cannot be
applied without further study of the Kullback distance-like function. Notice further that entropic-type
penalization terms in proximal algorithms have been the subject of an intensive research effort in
the mathematical programming community with the goal of handling positivity constraints; see \cite{Teboulle:1992}
and the references therein for instance. The analysis proposed here applies to the more general Kullback distance-like functions $I_y$ that occur in EM. Our goal is to show that such cluster points satisfy the
well known Karush-Kuhn-Tucker conditions of nonlinear programming which extend the stationarity condition $\nabla l_y(\theta)=0$ to the case where $\theta$ is subject to constraints. As before, it is straightforward 
to extend the proposed analysis to the case of penalized likelihood estimation. 

In the sequel, the distance-like function will be assumed to have the following additional properties.
\begin{ass}
\label{ass4}
The Kullback distance-like function $I_y$ is of the form
$$
I_y(\theta,\bar{\theta})=\sum_{1\leq i\leq n, 1\leq j\leq m} \alpha_{ij}(y_j)t_{ij}(\theta)
\phi\Big(\frac{t_{ij}(\bar{\theta})}{t_{ij}(\theta)}\Big),
$$
where for all $i$ and $j$, $t_{ij}$ is continuously differentiable on its domain of definition,
$\alpha_{ij}$ is a function from $\mathcal Y$ to $\mathbb R_+$, the set of positive real
numbers, and the function $\phi$ is a non negative convex continuously differentiable function defined for positive
real numbers only and such that $\phi(\tau)=0$ if and only if $\tau=1$.
\end{ass}
If $t_{ij}(\theta)=\theta_i$ and $\alpha_{ij}=1$ for all $i$ and all $j$, the function
$I_y$ is the well known $\phi$ divergence defined by Csisz\`ar in \cite{Csiszar:67}.
Assumption \ref{ass4} is satisfied in most standard examples (for instance Gaussian 
mixtures and Poisson inverse problems) with the choice
$\phi(\tau)=\tau\log(\tau)$. 

\subsection{More properties of the Kullback distance-like function}

The main property that will be needed in the sequel is that under Assumption \ref{ass4}, the function $I_y$ satisfies
the same property as the one given in Lemma \ref{asymreg} above, even on the boundary of its domain $D_I$. This is the result
of Proposition \ref{nondegphi} below. We begin with one elementary lemma.

\begin{lem}
\label{convone}
Under Assumptions \ref{ass4}, the function $\phi$ is decreasing on $(0,1)$, is increasing on $(1,+\infty)$ and $\phi(\tau)$
converges to $+\infty$ when $\tau$ converges to $+\infty$.
We have $\lim_{k\rightarrow +\infty} \phi(\tau^k)=0$ if and only if $\lim_{k\rightarrow +\infty}\tau^k=1$.
\end{lem}

{\bf Proof}. The first statement is obvious. For the second statement, the "if" part is trivial, so we only prove the "only if" part.
First notice that the sequence $(\tau^k)_{k\in \mathbb N}$ must be
bounded. Indeed, the level set $\{ \tau \mid \phi(\tau)\leq \gamma\}$ is bounded for all
$\gamma\geq 0$ and contains the sequence $(\tau^k)_{k\geq K}$ for $K$ sufficiently large. Thus,
the Bolzano-Weierstass theorem applies. Let $\tau^*$ be an accumulation point of
$(\tau^k)_{k\in \mathbb N}$. Since $\phi$ is continuous, we get that $\phi(\tau^*)=0$ and thus
we obtain $\tau^*=1$. From this, we deduce that the sequence has only one cluster point, which
is equal to 1. Therefore, $\lim_{k\rightarrow +\infty}\tau^k=1$.
\hfill$\Box$

Using these lemmas, we are now in position to state and prove the main property of $I_y$.
\begin{prop}
\label{nondegphi}
The following statements hold.

(i) For any sequence $(\theta^k)_{k\in \mathbb N}$ in $\mathbb R_+$ and any bounded sequence
$(\eta^k)_{k\in \mathbb N}$ in $\mathbb R_+$, the fact that 
$\lim_{k\rightarrow +\infty} I_y(\eta^k,\theta^k)=0$ implies
$\lim_{k\rightarrow +\infty} |t_{ij}(\eta^k)-t_{ij}(\theta^k)|=0$ for all $i$,$j$
such that $\alpha_{ij}\neq 0$.

(ii) If one coordinate of one of the two sequences $(\theta^k)_{k\in \mathbb N}$ and $(\eta^k)_{k\in \mathbb N}$ 
tends to infinity, so does the other's same coordinate.
\end{prop}

{\bf Proof}. Fix $i$ in $\{1,\ldots,n\}$ and $j$ in $\{1,\ldots,m\}$ and assume that
$\alpha_{ij}\neq 0$.

(i) We first assume that $(t_{ij}(\eta_i^k))_{k\in \mathbb N}$ is bounded
away from zero. 

Since  $\lim_{k\rightarrow +\infty} I_y(\theta^k,\eta^k)=0$, then
$\lim_{k\rightarrow +\infty} \phi(t_{ij}(\theta^k)/t_{ij}(\eta^k))=0$ and Lemma \ref{convone}
implies that $\lim_{k\rightarrow +\infty} t_{ij}(\theta^k)/t_{ij}(\eta^k)=1$. Thus,
$\lim_{k\rightarrow +\infty}(t_{ij}(\theta^k)-t_{ij}(\eta^k))/t_{ij}(\eta^k)=0$ and since
$t$ is continuous, $t_{ij}(\eta^k)$ is bounded. This implies that
$\lim_{k\rightarrow +\infty}|t_{ij}(\theta^k)-t_{ij}(\eta^k)|=0$.

Next, consider the case of a subsequence $(t_{ij}(\eta^{\sigma(k)}))_{k\in \mathbb N}$
which tends towards zero. For contradiction, assume the existence of a subsequence
$(t_{ij}(\theta^{\sigma(\gamma(k))})_{k\in \mathbb N}$ which remains bounded away from zero, i.e.
there exists $a>0$ such that $t_{ij}(\theta^{\sigma(\gamma(k))})_{k\in \mathbb N}\geq a$
for $k$ sufficiently large. Thus, for $k$ sufficiently large we get
$$
\frac{t_{ij}(\theta^{\sigma(\gamma(k))})}{t_{ij}(\eta^{\sigma(\gamma(k))})}\geq \frac{a}{t_{ij}(\eta^{\sigma(\gamma(k))})}>1,
$$
and due to the fact that $\phi$ is increasing on $(1,+\infty)$, we obtain
\begin{equation}
\label{toto}
t_{ij}(\eta^{\sigma(\gamma(k))}) \phi\Big(\frac{t_{ij}(\theta^{\sigma(\gamma(k))})}{t_{ij}(\eta^{\sigma(\gamma(k))})}\Big)\geq
t_{ij}(\eta^{\sigma(\gamma(k))}) \phi\Big(\frac{a}{t_{ij}(\eta^{\sigma(\gamma(k))})}\Big).
\end{equation}
On the other hand, Lemma \ref{convone} says that for any $b>1$, $\phi^\prime(b)>0$. Since $\phi$
is convex, we get
$$
\phi(\tau)\geq \phi(b)+\phi^\prime(b)(\tau-b).
$$
Take $\tau=a/t_{ij}(\eta^k)$ in this last expression and combine with (\ref{toto}) to obtain
$$
t_{ij}(\eta^{\sigma(\gamma(k))}) \phi\Big(\frac{t_{ij}(\theta^{\sigma(\gamma(k))})}{t_{ij}(\eta^{\sigma(\gamma(k))})}\Big)\geq
t_{ij}(\eta^{\sigma(\gamma(k))}) (\phi(b)+\phi^\prime(b)\Big(\frac{a}{t_{ij}(\eta^{\sigma(\gamma(k))})}-b \Big).
$$
Passing to the limit, we obtain
$$
0=\lim_{k\rightarrow +\infty}
t_{ij}(\eta^{\sigma(\gamma(k))}) \phi\Big(\frac{t_{ij}(\theta^{\sigma(\gamma(k))})}{t_{ij}(\eta^{\sigma(\gamma(k))})}\Big)\geq
a\phi^\prime(b)>0,
$$
which gives the required contradiction.

(ii) If $(t_{ij}(\theta^k))_{k\in \mathbb N}\rightarrow +\infty$ then 
$(t_{ij}(\eta^k))_{k\in \mathbb N}\rightarrow +\infty$ is a direct consequence
of part (i). Indeed, if $t_{ij}(\eta^k)$ remains bounded, part (i) says that
$\lim_{k\rightarrow +\infty} |t_{ij}(\eta^k)-t_{ij}(\theta^k)|=0$, which contradicts
divergence of $(t_{ij}(\theta^k))_{k\in \mathbb N}$.

Now, consider the case where $(t_{ij}(\eta^k))_{k\in \mathbb N}\rightarrow +\infty$. Then, a contradiction
is easily obtained if we assume that at least a subsequence $(t_{ij}(\theta^{\sigma(k)})_{k\in \mathbb N}$
stays bounded from above. Indeed, in such a case, we have
$$
\lim_{k\rightarrow +\infty} \frac{t_{ij}(\theta^{\sigma(k)})}{t_{ij}(\eta^{\sigma(k)})}=0,
$$
and thus, $\phi(t_{ij}(\theta^k)/t_{ij}(\eta^k))\geq \gamma$ for some $\gamma >0$ since
we know that $\phi$ is decreasing on $(0,1)$ and $\phi(1)=0$. This implies that
$$\lim_{k\rightarrow +\infty}t_{ij}(\eta^{\sigma(k)}) \phi\Big(\frac{t_{ij}(\theta^{\sigma(k)})}{t_{ij}(\eta^{\sigma(k)})}\Big)
=+\infty,$$
which is the required contradiction.
\hfill$\Box$

\subsection{Cluster points are KKT points}

The main result of this section is the property that any cluster point $\theta^*$ such that
$(\theta^*,\theta^*)$ lies on the boundary of $D_I$ satisfies the Karush-Kuhn-Tucker necessary
conditions for optimality on the domain of the log-likelihood function. In the context of Assumptions \ref{ass4}, $D_I$
is the set
$$
D_I=\{\theta \in \mathbb R^n \mid t_{ij}(\theta)>0 \hspace{.3cm} \forall i\in \{1,\ldots,n\} \text{ and } j
\in \{1,\ldots,m\}  \}.
$$

We have the following theorem.

\begin{thm}
\label{bord}
Let $\theta^*$ be a cluster point of the Kullback-proximal sequence. Assume that all the functions $t_{ij}$ are differentiable at $\theta^*$.
Let $\mathcal I^*$ be the set of all couples of indices
$(i,j)$ such that the constraint $t_{ij}(\theta)\geq 0$ is active at $\theta^*$, i.e. $t_{ij}(\theta^*)=0$. If
$\theta^*$ lies in the interior of $D_l$, then $\theta^*$ satisfies the Karush-Kuhn-Tucker necessary conditions for optimality, i.e. there
exists a family of reals $\lambda_{ij}$, $(i,j)\in \mathcal I^*$ such that
$$
\nabla l_y(\theta^*)+\sum_{(i,j)\in \mathcal I^*} \lambda_{ij} \nabla t_{ij}(\theta^*)=0.
$$
\end{thm}

{\bf Proof}. Let $\Phi_{ij}(\theta,\bar{\theta})$ denote the bivariate function defined by
$$\Phi_{ij}(\theta,\bar{\theta})= \phi\Big(\frac{t_{ij}(\bar{\theta})}{t_{ij}(\theta)}\Big).$$
Let $\{\theta^{\sigma(k)}\}_{k\in \mathbb N}$ be a
convergent subsequence of the proximal sequence with limit equal to $\theta^*$.
The first order optimality condition at iteration $k$ is given by
\begin{equation}
\label{frst}
\begin{array}{rl}
\nabla l_y(\theta^{\sigma(k)}) & +\beta_{\sigma(k)}\Big(\sum_{ij} \alpha_{ij}(y_j)\nabla t_{ij}(\theta^{\sigma(k)})
\phi\Big(\frac{t_{ij}(\theta^{\sigma(k)-1})}{t_{ij}(\theta^{\sigma(k)})}\Big) \\
& +\sum_{ij} \alpha_{ij}(y_j) t_{ij}(\theta^{\sigma(k)})
\nabla_1 \Phi(\theta^{\sigma(k)},\theta^{\sigma(k)-1})\Big)=0.
\end{array}
\end{equation}
We have
$$t_{ij}(\theta^{\sigma(k)})\nabla_1 \Phi(\theta^{\sigma(k)},\theta^{\sigma(k)-1})
=-\frac{t_{ij}(\theta^{\sigma(k)-1})}{t_{ij}(\theta^{\sigma(k)})}
\phi^{\prime}\Big( \frac{t_{ij}(\theta^{\sigma(k)-1})}{t_{ij}(\theta^{\sigma(k)})} \Big)
\nabla t_{ij}(\theta^{\sigma(k)})$$
for all $i$ and $j$. 

{\bf Claim A}.  {\em For all $(i,j)$ such that $\alpha_{ij}(y_j)\neq 0$, we have
$$
\lim_{k\rightarrow +\infty}t_{ij}(\theta^{\sigma(k)})\nabla_1 \Phi(\theta^{\sigma(k)},\theta^{\sigma(k)-1})=0.
$$
}

{\bf Proof of Claim A}. Two cases may occur. In the first case, we have $t_{ij}(\theta^*)=0$. Since
the sequence $\{\theta^k\}_{k\in \mathbb N}$ is bounded due to Lemma \ref{boundu}, continuous differentiability of $\phi$ and
the $t_{ij}$ proves that $\nabla_1 \Phi(\theta^{\sigma(k)},\theta^{\sigma(k)-1})$ is bounded from above. Thus, the desired conclusion follows.
In the second case, $t_{ij}(\theta^*)\neq 0$ and applying Lemma \ref{yal}, we deduce that
$I_y(\theta^{\sigma(k)},\theta^{\sigma(k)-1})$ tends to zero. Hence,
$\lim_{k\rightarrow +\infty} \Phi(\theta^{\sigma(k)},\theta^{\sigma(k)-1})=0$,
which implies that $\lim_{k\rightarrow +\infty}\theta^{\sigma(k)}/\theta^{\sigma(k)-1}=1$. From this
and Assumptions \ref{ass4}, we deduce that
$\lim_{k\rightarrow +\infty} \phi^{\prime}(t_{ij}(\theta^{\sigma(k)-1})/t_{ij}(\theta^{\sigma(k)}))=0$.
Since $\{\theta^{\sigma(k)}\}_{k\in \mathbb N}$ converges to $\theta^*$ and that $t_{ij}(\theta^*)\neq 0$,
we obtain that the subsequence $\{t_{ij}(\theta^{\sigma(k)-1})/t_{ij}(\theta^{\sigma(k)})\}_{k\in \mathbb N}$ is bounded from above.
Moreover, $\{\nabla t_{ij}(\theta^{\sigma(k)})\}_{k\in \mathbb N}$ is also bounded by continuous
differentiability of $t_{ij}$. Therefore, the fact that
$\lim_{k\rightarrow +\infty} \phi^{\prime}(t_{ij}(\theta^{\sigma(k)-1})/t_{ij}(\theta^{\sigma(k)}))=0$ establishes Claim A.
\hfill$\Box$

Using this claim, we just have to study the remaining right hand side terms in (\ref{frst}), namely the expression
$\sum_{ij} \alpha_{ij}(y_j)\nabla t_{ij}(\theta^{\sigma(k)}) \phi\Big(\frac{t_{ij}(\theta^{\sigma(k)-1})}{t_{ij}(\theta^{\sigma(k)})}\Big)$. 
Let $\mathcal I^{**}$ be a subset of the active indices $\mathcal I$ such that the family $\{\nabla t_{ij}(\theta^*)\}_{ij}$ is linearly 
independent. This linear independence is preserved under small perturbations, we may assume without 
loss of generality that the family $\Big\{\nabla t_{ij}(\theta^{\sigma(k)})\Big\}_{(i,j)\in \mathcal I^{**}}$ is linearly independent
for $k$ sufficiently large. For such $k$, we may rewrite equation (\ref{frst}) as 
\begin{equation}
\label{frstbis}
\begin{array}{rl}
\nabla l_y(\theta^{\sigma(k)}) & +\beta_{\sigma(k)}\Big(\sum_{(i,j)\in \mathcal I^{**}} \lambda^{\sigma(k)}_{ij}(y_j)\nabla t_{ij}(\theta^{\sigma(k)})\\
& +\sum_{ij} \alpha_{ij}(y_j) t_{ij}(\theta^{\sigma(k)})
\nabla_1 \Phi(\theta^{\sigma(k)},\theta^{\sigma(k)-1})\Big)=0.
\end{array}
\end{equation}

{\bf Claim B}. {\em The sequence $\{\lambda^{\sigma(k)}_{ij}(y_j)\}_{k\in \mathbb N}$ is bounded.}

{\bf Proof of claim B}. Using the previous claim and the continuous differentiability of $l_y$ and $t_{ij}$, equation (\ref{frstbis})
expresses that $\{\lambda^{\sigma(k)}_{ij}(y_j)\}_{ij}$ are proportional to the
coordinates of the projection on the span of the $\{\nabla t_{ij}(\theta^{\sigma(k)})\}_{ij}$ of a vector
converging towards $\nabla l_y(\theta^*)$. Since $\{\nabla t_{ij}(\theta^{\sigma(k)})\}_{ij}$, for $(i,j)\in\mathcal I^{**}$, form a linearly independent family
for $k$ sufficiently large, none of the coordinates can tend towards infinity. \hfill $\Box$

We are now in position to finish the proof of the theorem. Take any cluster point $\tau_{ij}$ of
$t_{ij}(\theta^{\sigma(k)-1})/t_{ij}(\theta^{\sigma(k)})$. Using Claim B, we know that 
$(\lambda^{\sigma(k)}_{ij}(y_j))_{(i,j)\in \mathcal I^{**}}$ lies in a compact set. Let 
$(\lambda^*_{ij})_{(i,j)\in \mathcal I^{**}}$ be a cluster point of this sequence. Passing to the limit, we obtain from equation (\ref{frst}) that
$$
\nabla l_y(\theta^{\sigma(k)})+\beta^*\Big(\sum_{(i,j)\in \mathcal I^{**}} \lambda_{ij}^* \nabla t_{ij}(\theta^*)\Big)=0.
$$
for every cluster point $\beta^*$ of $\{\beta_{\sigma(k)}\}_{k\in \mathbb N}$.
For all $(i,j) \in \mathcal I^{**}$, set $\lambda_{ij}=\beta^* \lambda^*_{ij}$. 
This equation is exactly the Karuch-Kuhn-Tucker
necessary condition for optimality. \hfill$\Box$

\begin{rem}
If the family $(\nabla t_{ij}(\theta^{\sigma(k)}))_{(i,j)\in \mathcal I^*}$ is linearly independent for $k$ 
sufficiently large, Theorem \ref{bord} holds and in addition the $\{\lambda_{ij}\}_{ij}$ are 
nonnegative, which proves that $\theta^*$ satisfies the Karush-Kuhn-Tucker conditions when it lies 
in the closure of $\mathcal D_I$.  
\end{rem}

\section{Application}

The goal of this section is to illustrate the utility of the previous theory for a nonparametric survival analysis 
with competing risks proposed by Ahn, Kodell and Moon in \cite{Ahn&al:00}.   

\subsection{The problem and the Kullback proximal method}
\label{prob2}
This problem can be described as follows. Consider 
a group of $N$ animals in an animal carcinogenecity experiment. 
Sacrifices are performed at certain prescribed times denoted 
by $t_1,t_2, \ldots, t_m$ in order to study the presence of the tumor of interest. Let $T_1$ be the time to onset of
tumor, $T_D$ the time to death from this tumor and $X_C$ be the 
time to death from a cause other than this tumor. Notice that $T_1$, $T_D$ and $X_C$ are unobservable. 
The quantities to be estimated  are $S(t)$, $P(t)$ and 
$Q(t)$, the survival function of  $T_1$, $T_D$ and $X_C$ respectively. It is assumed that $T_1$ and $T_D$ are 
statistically independent of $X_C$.

A nonparametric approach to estimation of $S$, $P$ and $Q$ is proposed in \cite{Ahn&al:00}: observed data $y_1,\ldots,y_n$ are the number of deaths on every interval $(t_j,t_{j+1}]$ 
which can be classified into the following four categories,  
\begin{itemize}
\item death with tumor (without knowing cause of death)
\item death without tumor
\item sacrifice with tumor 
\item sacrifice without tumor
\end{itemize}
This gives rise to a multinomial model whose probability mass is parametrized by the values of $S$, $P$ and $Q$ at times 
$t_1,\ldots, t_m$. More precisely, for each time interval $(t_j,t_{j+1}]$ denote by $c_j$ the number of deaths with 
tumor present, $b_{1j}$ the number of deaths with tumor absent, $a_{2j}$ the number of sacrifices with tumor present
and $b_{2j}$ the number of sacrifices with tumor absent. Let $N_j\leq N$ be the number of live animals in the population at $t_j$, it is shown in \cite{Ahn&al:00} that the corresponding log-likelihood is given by 
\begin{equation}
\begin{array}{rl}
\log g(y;\theta) & =\sum_{j=1}^m (N_{j-1}-N_j)\sum_{k=1}^{j-1} \log(p_kq_k) + (a_{2j}+b_{2j}) \log(p_jq_j) \\
& + c_j \log\Big((1-p_j)+(1-\pi_jp_j)(1-q_j) \Big) \\
& + b_{1j}\log ((1-q_j)\pi_{j-1}) +a_{2j} \log (1-\pi_j) + b_{2j} \log \pi_j + Cst,
\end{array}
\end{equation}
where $Cst$ is a constant $\pi_j=S(t_j)/P(t_j)$, $p_j=P(t_j)/P(t_{j-1})$ and $q_j=Q(t_j)/Q(t_{j-1})$, $j=1,\ldots,m$, $\theta=(\pi_1,\ldots,p_J,p_1,\ldots,p_J,q_1,\ldots,q_J)$ and the parameter space is specified by the constraints 
\begin{equation}
\begin{array}{rl}
\Theta = & \Big\{ \theta = (\pi_1,\ldots,p_J,p_1,\ldots,p_J,q_1,\ldots,q_J) \mid 
0\leq \pi_j \leq 1,\\
& 0\leq p_j\leq 1, \hspace{.3cm} 0\leq q_j\leq 1, \hspace{.3cm} j=1,\ldots,m
\text{ and } \pi_j p_j \leq \pi_{j-1} \hspace{.3cm} j=2,\ldots,m \Big\},
\end{array}
\end{equation}
where the last nonconvex constraint serves to impose monotonicity of $S$. Note that monotonicity of $P$ and $Q$ is a direct consequence of the constraints on the $p_j$'s and the $q_j$'s, respectively. 

Define the complete data $x_1,\ldots,x_n$ as a measurement that indicates the cause of death in addition 
to the presence of absence of a tumor in the dead animals. Specifically, $x_1,\ldots,x_n$ 
should fall into one of the following categories 
\begin{itemize}
\item death caused by tumor and death with incidental tumor
\item death without tumor
\item sacrifice with tumor 
\item sacrifice without tumor
\end{itemize}
To each time interval $(t_j,t_{j+1}]$ among those animals dying of natural causes, there correspond 
the numbers $d_j$ of deaths caused by tumor and the number $a_{1j}$ of deaths with incidental tumor, neither of which are observable. The associated complete log-likelihood function is given by 
\begin{equation}
\begin{array}{rl}
\log f(x;\theta) & =\sum_{j=1}^m (N_{j-1}-N_j)\sum_{k=1}^{j-1} \log(p_kq_k) + (a_{2j}+b_{2j}) \log(p_jq_j) \\
& + d_j \log (1-p_j) + a_{1j} \log \Big((1-\pi_jp_j)(1-q_j) \Big) \\
& + b_{1j}\log ((1-q_j)\pi_{j-1}) +a_{2j} \log (1-\pi_j) + b_{2j} \log \pi_j + Cst
\end{array}
\end{equation}
Now, we have to compute the expectation $Q(\theta,\bar{\theta})$ of the log-likelihood function of the complete data  conditionally to the parameter $\bar{\theta}$. The random variables $d_j$ and $a_{1j}$ are binomial with parameter $\lambda_j$ and $1-\lambda_j$ where $\lambda_j$ is the probability that the death was caused by the tumor conditioned on the presence of the tumor. Conditioned on $\bar{\theta}$, we have 
\begin{equation}
\lambda_j=\frac{1-\bar{p}_j}{1-\bar{p}_j+(1-\bar{\pi}_j \bar{p}_j)(1-\bar{q}_j)}
\end{equation}  
(see \cite[Section 3]{Ahn&al:00} for details). From this, we obtain that the conditional mean values of $d_j$ and $a_{1j}$ are given by 
\begin{equation}
{\rm E}[d_j \mid y;\bar{\theta}]=\lambda_j c_j \hspace{.4cm} \text{ and }\hspace{.4cm} {\rm E}[a_{1j} \mid y;\bar{\theta}]=(1-\lambda_j) c_j. 
\end{equation}
Therefore
\begin{equation}
\begin{array}{rl}
Q(\theta,\bar{\theta}) & =\sum_{j=1}^m (N_{j-1}-N_j)\sum_{k=1}^{j-1} \log(p_kq_k) + (a_{2j}+b_{2j}) \log(p_jq_j) \\
& + \lambda_j c_j\log (1-p_j) + (1-\lambda_j) c_j \log \Big((1-\pi_jp_j)(1-q_j) \Big) \\
& + b_{1j}\log ((1-q_j)\pi_{j-1}) +a_{2j} \log (1-\pi_j) + b_{2j} \log \pi_j + Cst. 
\end{array}
\end{equation}
From this, we can easily compute the associated Kullback distance-like function:
\begin{equation}
I_y(\theta,\bar{\theta})= \sum_{j=1}^m c_j \Big(t_j^\prime(\theta) \phi\Big(\frac{t_j^\prime(\bar{\theta})}{t_j^\prime(\theta)} \Big)+
 t_j^{\prime\prime}(\theta) \phi\Big(\frac{t_j^{\prime\prime}(\bar{\theta})}{t_j^{\prime\prime}(\theta)} \Big)\Big),
\end{equation}
with 
\begin{equation}
t_j^\prime(\theta)=\frac{1-p_j}{1-p_j+(1-\pi_jp_j)(1-q_j)} \hspace{.4cm} \text{ and } \hspace{.4cm}t_j^{\prime\prime}(\theta)=\frac{(1-\pi_jp_j)(1-q_j)}{1-p_j+(1-\pi_jp_j)(1-q_j)}
\end{equation}
and $\phi$ is defined by $\phi(\tau)=\tau \log (\tau)$. It is straightforward to verify that   
Assumptions \ref{ass1}, \ref{ass2}, \ref{ass3} and \ref{ass4} are satisfied. 

The main computational problem in this example is to handle the difficult nonconvex constraints entering the definition 
of the parameter space $\Theta$. The authors of \cite{Moon&al:99} and \cite{Ahn&al:00} use the Complex Method proposed by  Box in \cite{Box:65} to address this problem. However, the theoretical convergence properties of Box's method are not 
known as reported in article MR0184734 in the Math. Reviews. Using our proximal point framework, we are able to easily incorporate the nonconvex constraints into the Kullback distance-like function and obtain an efficient algorithm with 
satisfactory convergence properties. For this purpose, let $I_y^\prime$ be defined by 
\begin{equation}
I_y^\prime(\theta,\bar{\theta})=I_y(\theta,\bar{\theta})+ \sum_{j=2}^m t_j^{\prime\prime\prime}(\theta) \phi\Big(\frac{t_j^{\prime\prime\prime}(\bar{\theta})}{t_j^{\prime\prime\prime}(\theta)}\Big) 
\end{equation}  
where 
\begin{equation}
t_j^{\prime\prime\prime} (\theta)=\frac{\pi_{j-1}-\pi_jp_j}{\sum_{i=2}^m \pi_{i-1}-\pi_ip_i}. 
\end{equation}
Using this new function, the nonconvex constraints $\pi_jp_j\leq \pi_{j-1}$ are satisfied for all proximal iterations and Assumptions \ref{ass4} still hold. 

\subsection{Experimental results}
We implemented the Kullback proximal algorithm with different choices of relaxation sequence 
$(\beta_k)_{k\in \mathbb N}$, $\beta_k=\beta$. The M-step of the EM algorithm does not have a 
closed form solution, so that nothing is lost by setting $\beta_k$ to a constant not equal to one. 

We attempted to supplement the KPP-EM algorithm with the Newton method and other built-in methods available 
in Scilab but they were not even able to 
find local maximizers due to the explosive nature of the logarithms near zero, leading 
these routines to repetitive crashes. To overcome this difficulty, we found it convenient to use the extremely simple simulated annealing random 
search procedure; see \cite{Zabinsky:03} for instance. This random search approach avoids numerical difficulties 
encountered using standard optimization packages and easily handles nonconvex
constraints. The a.s. convergence of this procedure 
is well established and recent studies such as \cite{Kalai:06} confirm the good computational efficiency for convex functions optimization.  

Some of our results for the data of Table 1 of \cite{Moon&al:99} are given in Figures 1 to 4.
In the reported experiments, we chose three constant sequences with respective values $\beta_n=100,\: 1, \: .01$. 
We observed the following phenomena

\vspace{.3cm}

{\bf 1.} after one hundred iterations the increase in the likelihood function is less than $10^{-5}$ except for 
the case $\beta_n=100$ (Figure \ref{fig4}) where the algorithm had not converged. 

{\bf 2.} for $\beta_n=100$ we often obtained the best initial growth of the likelihood  

{\bf 3.} for $\beta_n=.01$ we always obtained the highest likelihood when the number of iterations was limited to 
50 (see Figure \ref{fig3} for the case MCL Male AL).

\vspace{.3cm}

It was shown in \cite{Chretien&Hero:00} that penalizing with a parameter sequence 
$(\beta_n)_{n \in \mathbb N}$ converging towards zero implies superlinear convergence in the case where 
the maximum likelihood estimator lies in the interior of the constraint set. Thus, our simulations 
results seem to confirm observation 3. The second observation was surprising to us but this phenomenon occured repeatedly in our experiments. This behavior did not occur in our simulations for the Poisson inverse problem in 
\cite{Chretien&Hero:00} for instance.

In conclusion, this competing risks estimation problem is an interesting test for our Kullback-proximal method which 
shows that the proposed framework can provide provably convergent methods for difficult constrained nonconvex estimation problems for which standard optimization algorithms can be hard to tune. The relaxation parameter sequence $(\beta_n)_{n\in \mathbb N}$ also appeared crucial for this problem although 
the choice $\beta_n=1$ could not really be considered unsatisfactory in practice.

\begin{figure}[hbtp]
\begin{center}

\includegraphics[angle=-90]{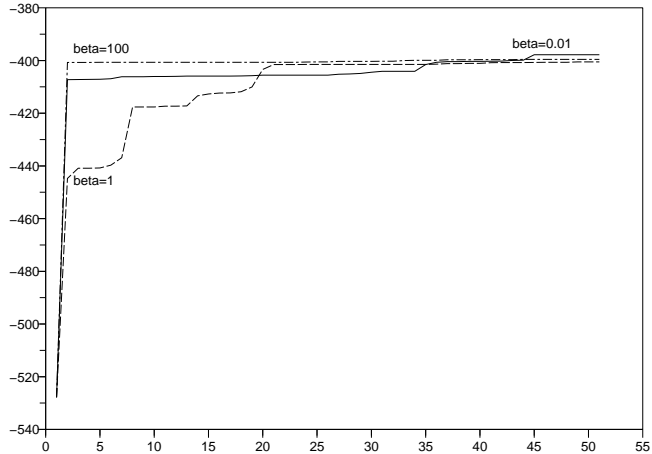}

\end{center}
\caption{\label{fig1} Evolution of the log-likelihood versus iteration number: MCL Female CR case}
\end{figure}

\begin{figure}[hbtp]
\begin{center}

\includegraphics[angle=-90]{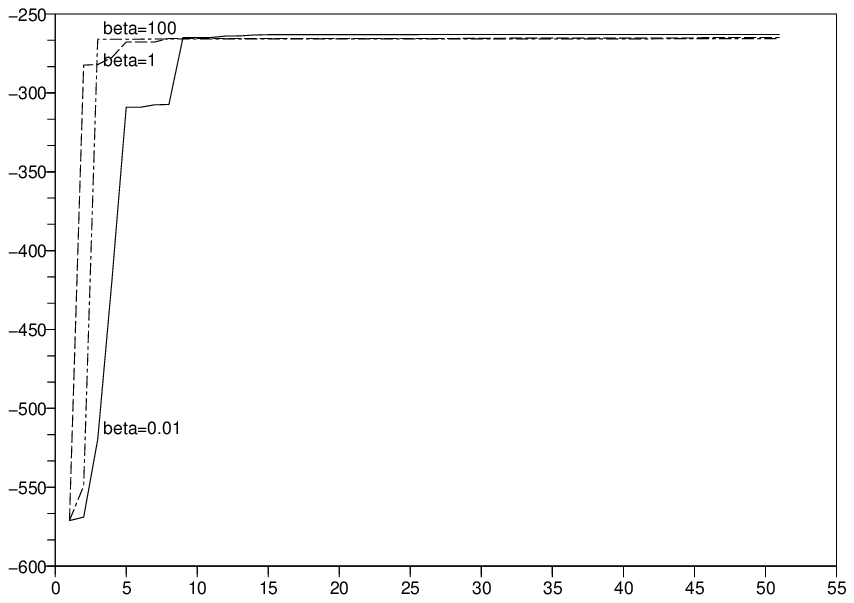}

\end{center}
\caption{\label{fig2} Evolution of the log-likelihood versus iteration number: MCL Male AL case}
\end{figure}

\begin{figure}[hbtp]
\begin{center}

\includegraphics[angle=-90]{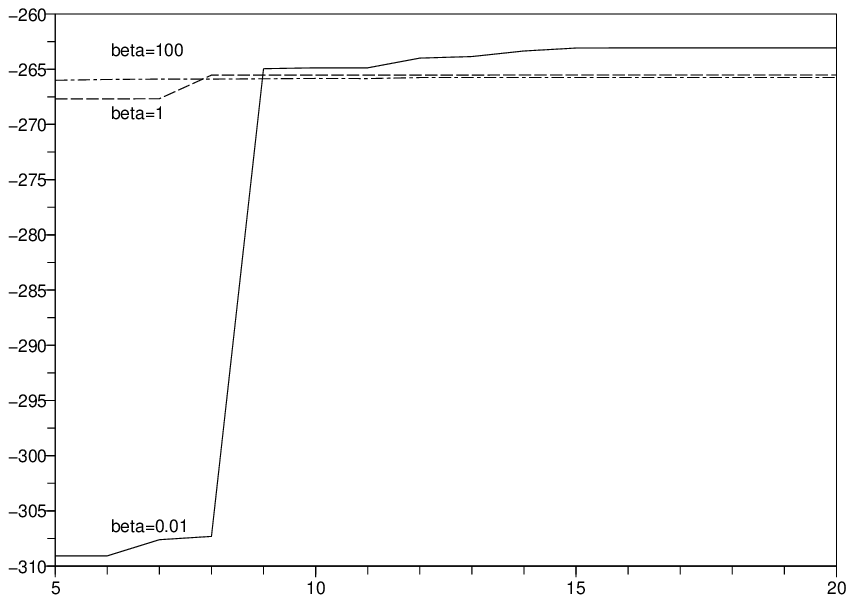}

\end{center}
\caption{\label{fig3} Evolution of the log-likelihood versus iteration number: Detail of MCL Male AL case}
\end{figure}

\begin{figure}[hbtp]
\begin{center}

\includegraphics[angle=0]{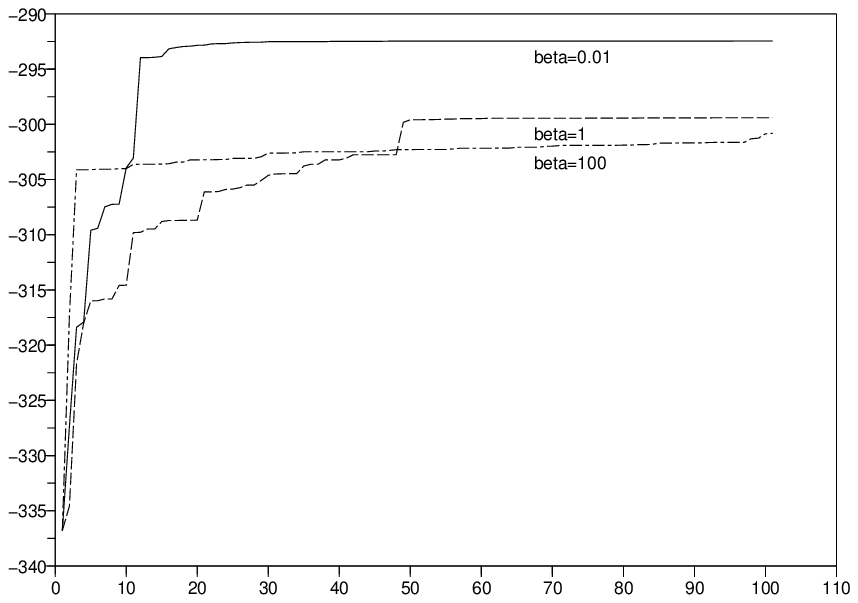}

\end{center}
\caption{\label{fig4} Evolution of the log-likelihood versus iteration number: MCL Female AL case}
\end{figure}

\section{Conclusions}

The goal of this paper was the study of the asymptotic behavior of the EM algorithm and its proximal generalizations. 
We clarified the analysis by making use of the Kullback-proximal theoretical framework. Two of our main contributions 
are the following. Firstly we showed that interior cluster points are stationary points of the likelihood function and are local maximizers for sufficiently small values of $\beta$. Secondly, we showed that cluster points lying on the boundary satisfy the Karush-Kuhn-Tucker conditions.
Such cases were very seldom studied in the literature although constrained estimation is a topic of growing importance; see for instance the special issue of the Journal of Statistical Planning and Inference \cite{jspi107:02} which is devoted to the problem of estimation under constraints. On the negative side, the analysis from the Kullback-proximal viewpoint allowed us to understand why uniqueness of the cluster point is hard to establish theoretically. On the positive side, we were able to implement a new and efficient proximal point method for estimation in the difficult tumor lethality problem  involving nonlinear inequality constraints.

\end{document}